\documentclass[12pt]{article}

\usepackage{amsmath,amssymb,graphicx,jheppub, comment, braket}

\usepackage{tikz}
\usepackage{amsmath,latexsym}
\usepackage{epsfig,amssymb,euscript,xspace, color}
\usepackage{amsmath,mathtools,empheq,amsthm}
\usepackage{graphicx, verbatim, cancel}
\usepackage[most]{tcolorbox}
\usepackage{annotate-equations}
\usepackage{esint}
\usepackage{empheq}
\usepackage{mathrsfs} 
\usepackage{appendix}

\title{ \Large Celestial dual of conformal gravity MHV amplitudes: \\ an OPE analysis}

\author{Nirmal Ghorai, Partha Paul and Nemani V. Suryanarayana}
\affiliation[]{The Institute of Mathematical Sciences, \\ IV Cross Road, CIT Campus, Taramani, Chennai 600113, India. \\ \&}
\affiliation[]{Homi Bhabha National Institute, \\ Training School Complex, Anushakti Nagar, Mumbai 400094, India. \\ }
\emailAdd{nirmalg@imsc.res.in}
\emailAdd{parthapaul@imsc.res.in}
\emailAdd{nemani@imsc.res.in}

\abstract{In an earlier paper [\href{https://arxiv.org/abs/2511.03669}{arXiv:2511.03669}] we extracted the OPE of celestial CFT operator duals of positive helicity graviton and scalar particles from the Mellin transformed relevant MHV amplitudes of conformal gravity, realised as the bosonic subsector of the Berkovits--Witten theory. 
A soft theorem analysis of bulk MHV amplitudes established that this conformal 
gravity exhibits a chiral $\mathfrak{bms}_4$ symmetry on the celestial sphere with the associated $\mathfrak{sl}(2,\mathbb{R})$ current algebra, which acquires a non-trivial central extension, unlike the Einstein gravity. Here we construct a $2d$ chiral CFT free-field 
realisation of the relevant chiral $\mathfrak{bms}_4$ algebra in terms of three free scalars (\(\phi_i\))
and three $(\beta_i,\gamma_i)$ ghost pairs, and propose vertex operators for the 
positive-helicity graviton primary $G^{++}_{\Delta}(z,\bar{z})$ as well as the scalar primary  $\Phi_{\Delta}(z,\bar{z})$, and compute their 
OPEs. These OPEs reproduce exactly those obtained from the bulk conformal gravity MHV amplitudes, providing a concrete 
celestial dual description of its MHV sector.}
\begin{document}
\maketitle

\section{Introduction}
 Motivated by the remarkable success of the AdS/CFT 
correspondence, considerable effort has been devoted to extending holographic ideas to asymptotically flat spacetimes. Celestial holography~\cite{Pasterski:2016qvg, Strominger:2017zoo, Pasterski:2017kqt, Banerjee:2018gce, 
Raclariu:2021zjz, Pasterski:2021rjz} has emerged as a promising avenue toward this goal. It proposes that the holographic dual of quantum gravity in four-dimensional (4$d$) asymptotically flat spacetime is a putative two-dimensional (2$d$) conformal field theory living on the celestial sphere at null infinity $\mathscr{I}^{+}$, known as celestial CFT. The holographic dictionary is very simple. A massless particle with energy $\omega$ and helicity $\sigma$ is represented by a $2d$ conformal primary operator under Lorentz ($\mathrm{SL}(2,\mathbb{C})$) transformations on the celestial sphere with dimension $\Delta$ and spin $\sigma$. This is achieved technically via a Mellin transform with respect to the particles' energies, while the momentum directions determine the insertion points of the corresponding primary operators on the celestial sphere \cite{Strominger:2017zoo, Pasterski:2017kqt, Banerjee:2018gce}. The correlation functions of these primary operators then encode the bulk $S$-matrix through the Mellin transform \cite{Strominger:2017zoo, Pasterski:2016qvg, Pasterski:2017kqt, Banerjee:2018gce,  Raclariu:2021zjz, Pasterski:2021rjz, Pasterski:2017ylz}.  This repackaging not only makes the  2$d$ conformal structure manifest, but also brings the well-developed machinery of CFT to bear on questions about flat-space scattering.

Holographic duality is most convincing when the symmetries on both sides match precisely. The central motivation behind celestial holography is the isomorphism between the Lorentz group and the group of global conformal transformations ($\mathrm{SL}(2,\mathbb{C})$) on the celestial sphere. Beyond this, soft graviton theorems imply that the global conformal symmetry on the celestial sphere is enhanced to an infinite-dimensional local  algebra~\cite{Strominger:2017zoo,Strominger:2013jfa,He:2014laa,Kapec:2014opa,Campiglia:2014yka,Avery:2015gxa,Campiglia:2015lxa,Avery:2015rga,Banerjee:2020zlg,Banerjee:2021dlm,Campiglia:2016efb,He:2017fsb, Campiglia:2016jdj}. In particular, in \cite{Banerjee:2020zlg}, Weinberg's soft graviton theorem~\cite{ Weinberg:1965kv} was reinterpreted as the Ward identities of two holomorphic supertranslation currents on the celestial sphere, while the subleading soft graviton theorem~\cite{Cachazo:2014fwa} was reinterpreted as the Ward identities of three holomorphic currents forming a chiral $\mathfrak{sl}(2,\mathbb{R})$ current algebra. This algebra, along with the holomorphic stress tensor, was subsequently identified as the asymptotic symmetry algebra of Einstein gravity in 4$d$ flat spacetime~\cite{Gupta:2021cwo}. We refer to this algebra as the chiral $\mathfrak{bms}_4$ algebra. Thus, any consistent celestial CFT dual to bulk Einstein gravity in $4d$ flat spacetime ought to possess (at least) the chiral $\mathfrak{bms}_4$ algebra as its underlying symmetry. A CFT is completely specified once the spectrum of primary operators and their operator product expansions (OPE)  are given.
Thus, one is naturally led to ask how the chiral $\mathfrak{bms}_4$ algebra is realised in the celestial CFT at the level of the OPE.  

In a celestial CFT, the OPE between two primary operators can be obtained in two complementary ways - either through symmetry-based arguments that fix the leading OPE coefficient, or by Mellin transforming the collinear singularities of bulk amplitudes. Celestial OPEs between different primary operators have been studied in the literature for various theories including  Einstein gravity, Yang-Mills theory, conformal gravity, etc. \cite{Banerjee:2020zlg,Banerjee:2021dlm,Pate:2019lpp,Guevara:2021abz,Strominger:2021lvk,Himwich:2021dau,Melton:2022fsf,Ball:2021tmb,Adamo:2021lrv,Costello:2022wso,Costello:2022upu,Costello:2023vyy,Ball:2023qim,Banerjee:2023zip,Banerjee:2020vnt,Ebert:2020nqf,Banerjee:2023rni,Adamo:2022wjo,Ren:2023trv,Bhardwaj:2022anh,Krishna:2023ukw,Hu:2022bpa,Banerjee:2023sdg,Ghorai:2025ebc}. By taking appropriate conformal soft limits at the level of the OPE involving positive helicity graviton primaries, one can further realise the chiral $\mathfrak{bms}_4$ algebra. Despite these successes, an intrinsically defined celestial CFT dual of gravitational theories - one whose correlation functions directly reproduce the bulk scattering amplitudes - remains elusive. In this paper, we take a step toward filling this gap by constructing a celestial dual for the MHV sector of conformal gravity. A step towards the construction of a dual description of $4d$ MHV gluon amplitudes in terms of correlators of $2d$ celestial CFT vertex operators has been proposed in \cite{Melton:2024akx}. This current work may be seen as an attempt at extending this line of research to bulk gravitational theories.

In particular, we focus on the Berkovits--Witten (BW) theory \cite{Berkovits:2004jj}, a superconformal gravity whose field content arises from a specific twistor-string theory. The tree-level scattering amplitudes of this theory have been studied in ~\cite{Berkovits:2004jj, Johansson:2017srf, Johansson:2018ues}. By focusing on a particular bosonic subsector of the BW theory, the authors of ~\cite{Johansson:2017srf, Johansson:2018ues} showed that the tree-level scattering amplitudes can be obtained from the double copy of two gauge theories. The gauge theories are (super-) Yang-Mills theory and a theory with a four-derivative kinetic term of the form $(DF)^2$. In~\cite{Ghorai:2025ebc}, this particular sector of the BW theory was considered by us. There, we showed that the leading and subleading soft graviton expansion of the tree-level MHV amplitudes of the BW theory still follow
as a consequence of the chiral supertranslations and chiral $\mathfrak{sl}(2,\mathbb{R})$ current algebra symmetries. However, somewhat interestingly, the realisation of the $\mathfrak{sl}(2,\mathbb{R})$ current algebra is quite different from the usual, and involves new representations that use particle changing operators. In addition, the OPE between two positive-helicity graviton primaries $G^{++}_{\Delta}(z,\bar{z})$ was derived, yielding~\cite{Ghorai:2025ebc},
\begin{equation}\label{gg_OPE_summary}
\begin{gathered}
G^{++}_{\Delta_1}(z,\bar z)\, G^{++}_{\Delta_2}(w,\bar w) = - \frac{(\bar z - \bar w)}{(z-w)}\,B(\Delta_1-1,\Delta_2-1)\,G^{++}_{\Delta_1+\Delta_2}(w,\bar w) \\
 - \frac{(\bar z- \bar w)^2}{(z-w)^2}\, B(\Delta_1,\Delta_2)\, \Phi_{\Delta_1+\Delta_2}(w,\bar w) + \cdots, \\
 \end{gathered}
\end{equation}
where $B(\Delta_1, \Delta_2)$ 
is the Euler beta function, and $\Phi_{\Delta}(z, \bar z)$ is a scalar primary, with the ellipsis representing the descendants. The OPE between a positive-helicity graviton primary $G^{++}_{\Delta}(z,\bar{z})$ and the scalar primary $\Phi_\Delta(z, \bar z)$ is \cite{Ghorai:2025ebc}
\begin{equation}
\label{com_graviton_scalar}
G^{++}_{\Delta_1}(z,\bar{z})\,\Phi_{\Delta_2}(w,\bar{w})
= -\frac{(\bar{z}-\bar{w})}{(z-w)}\,
B(\Delta_1-1,\Delta_2+1)\,\Phi_{\Delta_1+\Delta_2}(w,\bar{w}) + \cdots\, ,
\end{equation}
with the OPE between two scalar primaries being non-singular. 

Guided by these results, in this paper, we proceed to construct a candidate celestial CFT endowed with chiral $\mathfrak{bms}_4$ symmetry, and identify primary operators whose OPE structure precisely reproduces the behaviour \eqref{gg_OPE_summary} and \eqref{com_graviton_scalar} inferred from the bulk amplitude analysis.

We exploit the fact that the chiral $\mathfrak{bms}_4$ is a subalgebra of the chiral conformal $\mathfrak{bms}_4$ algebra we presented in \cite{Ghorai:2025ebc} and reviewed in Appendix~\ref{con_bms4}. This latter algebra is obtainable as the Drinfeld-Sokolov (DS) reduction of $\mathfrak{so}(2,4)$ (or the closely related $\mathfrak{sl}(4, \mathbb{R})$ ) current algebra. One way to achieve the DS reduction is
by implementing the BRST cohomology technique of de Boer et al. \cite{deBoer:1992sy, deBoer:1993iz}. We choose this approach as it directly provides a free-field realisation of the chiral $\mathfrak{bms}_4$ algebra and enables us to construct vertex operators in terms of these free fields transforming covariantly under this symmetry algebra. 
Then restricting attention to an appropriate subsector of these vertex operators yields OPEs consistent with the expected asymptotic symmetry algebra, namely the chiral $\mathfrak{bms}_4$. 

Each vertex operator is chosen to belong to a highest-weight representation of the $\mathfrak{sl}(2,\mathbb{R})_\kappa$ current algebra at level $\kappa$ \cite{DiFrancesco:1997nk, Polchinski:1998rq, Morozov:2025ngh, Creutzig:2020ffn}. By carefully analysing the holomorphic and anti-holomorphic conformal dimensions, we construct the candidate positive-helicity graviton primary $G^{++}_{\Delta}(z,\bar{z})$ as
\begin{equation}\label{intro_g_rep}
    G^{++}_{\Delta}(z,\bar z)
    = -\Delta\,\Gamma(\Delta-2) \sum_{n=0}^\infty \bar z^n\, U_{h, \bar h+n}(z) \;+\; \frac{\kappa+1}{\Delta+\kappa+1}\,\Gamma(\Delta) \sum_{n=0}^\infty \bar z^n\, V_{h, \bar h+n}(z),
\end{equation}
where $\Delta = h + \bar{h}$, the helicity is $\sigma = h - \bar{h} = 2$. The vertex operators  $U_{h,\bar{h}+n}(z)$ and $V_{h,\bar{h}+n}(z)$ belong to two distinct representations 
of the $\mathfrak{sl}(2,\mathbb{R})$ current algebra, of spin--$\tfrac{1}{2}$ and spin--$1$, respectively. 

The graviton primary faithfully reproduces the graviton-graviton OPE~\eqref{gg_OPE_summary},  including the emergence of the scalar primary $\Phi_{\Delta}(z,\bar{z})$, in 
exact agreement with the Mellin transformed bulk conformal gravity computation as in \cite{Ghorai:2025ebc}. Furthermore, the  $\mathfrak{sl}(2,\mathbb{R})$ current algebra acquires a non-trivial central extension, 
and we show that the correct level is $\kappa=-2$. Even though in the Einstein gravity case the $\mathfrak{sl}(2,\mathbb{R})$ current algebra is expected to be centerless, this is not the case in the conformal gravity context. Noticing the fact that the soft limit of the scalar operator $\Phi_\Delta$ is the identity operator, in turn gives the $\mathfrak{sl}(2,\mathbb{R})$ current algebra with this non-trivial center. This identification provides a concrete and self-consistent celestial dual description of the MHV sector of conformal gravity.

The rest of this paper is organised as follows.
In Section~\ref{free_field_rep}, we construct the free-field realisation of the chiral $\mathfrak{bms}_4$ algebra in terms of three free scalars (\(\phi_i\)) and three $(\beta_i,\gamma_i)$ ghost pairs, and present the resulting OPE algebra. In Section~\ref{spin_two_primary}, we build the highest-weight vertex operators and construct the spin-$2$ graviton primary $G^{++}_{\Delta}(z,\bar{z})$, fixing the appropriate combination using soft limit constraints. In Section~\ref{ope_two_graviton}, we compute the graviton--graviton OPE order by order.

Several technical details are collected in the appendices. Appendix~\ref{con_bms4} reviews the chiral conformal $\mathfrak{bms}_4$ algebra and its free-field realisation. In Appendices ~\ref{order_o2} \& \ref{order_o3}, we present details of the computation of the graviton-graviton OPE. In Appendix~\ref{symmetry_algebra}, we analyse the symmetry algebra by taking soft limits of the bulk graviton OPE and show that the level of the $\mathfrak{sl}(2,\mathbb{R})$ current algebra is $\kappa=-2$.
%
\section{A free-field realisation of the chiral 
\texorpdfstring{$\mathfrak{bms}_4$}{bms4} algebra}
\label{free_field_rep}
%
In this section, we construct a free-field realisation of the chiral  $\mathfrak{bms}_4$ algebra, with the aim of constructing a primary as a vertex operator with the correct quantum numbers to be a candidate graviton primary. The chiral $\mathfrak{bms}_4$ algebra is generated by six holomorphic currents: three $\mathfrak{sl}(2,\mathbb{R})$ currents $H^{0}_{a}(z)$ with $a = 0,\pm 1$ and conformal weights $(1,-a)$, two supertranslation currents $H^{1}_{\alpha}(z)$ with $\alpha = \pm\tfrac{1}{2}$ and weights $\bigl(\tfrac{3}{2},-\alpha\bigr)$, and a holomorphic stress tensor $T(z)$, with the OPEs given by:
\begin{equation}
\label{ope_one}
\begin{gathered}
T(z)\,T(w) \sim \frac{c/2}{(z-w)^4}
+\frac{2\,T(w)}{(z-w)^2}+\frac{\partial_w T(w)}{z-w}\,,\\ 
H^0_a(z)\,H^0_b(w) \sim \frac{-\frac{k}{2}\,\eta_{ab}}{(z-w)^2}
+\frac{f^{c}_{~ab}\,H^0_c(w)}{z-w}\,,\\[6pt]
T(z)\,H^0_a(w) \sim 
\frac{H^0_a(w)}{(z-w)^2}
+\frac{\partial_w H^0_a(w)}{z-w}\,, \\ 
T(z)\,H^1_\alpha(w) \sim 
\frac{\tfrac{3}{2}\,H^1_\alpha(w)}{(z-w)^2}
+\frac{\partial_w H^1_\alpha(w)}{z-w}\,,\qquad \\
H^0_a(z)\,H^1_\alpha(w) \sim 
\frac{(\lambda_a)^{\beta}_{\ \alpha}\,H^{1}_{\beta}(w)}{z-w}\,,\qquad
H^1_\alpha(z)\,H^1_\beta(w) \sim 0\,,\\[6pt]
\end{gathered}
\end{equation}
where $ \alpha, \beta \in \{\tfrac{1}{2}, -\tfrac{1}{2}\}$ and 
$a,b,c \in \{0, \pm 1\}$. The consistency of the chiral $\mathfrak{bms}_4$ algebra with OPE associativity does not restrict values of the Virasoro central charge ($c$) and the level of $\mathfrak{sl}(2,\mathbb{R})$ current algebra ($k$). However, we will use a specific free-field realisation in which they are related as follows 
\begin{equation}
\label{candk}
c = \frac{3(34+9\kappa)}{4+\kappa}\,, \qquad k = \kappa+1\, . \\[6pt]
\end{equation}
The non-vanishing components of the Killing metric $\eta_{ab}$, and the structure constants $f^c_{~ab}, \, (\lambda_a)^{\alpha}_{\ \beta}$, are given by
\begin{equation}\label{current_alg_coeff}
\begin{gathered}
\eta_{+-} = 2\,,\quad \eta_{00} = -4\,,\quad \eta_{-+} = 2\,,\\[4pt]
f^{-}_{-0} = -2\,,\quad f^{+}_{+0} = 2\,,\quad f^{0}_{+-} = 1\,,
\quad f^c_{ab} = -f^c_{ba}\,,\\[4pt]
(\lambda_+)^{\frac{1}{2}}_{\ -\frac{1}{2}} = 1\,,\quad
(\lambda_-)^{-\frac{1}{2}}_{\ \frac{1}{2}} = -1\,,\quad
(\lambda_0)^{\frac{1}{2}}_{\ \frac{1}{2}} = -1\,,\quad
(\lambda_0)^{-\frac{1}{2}}_{\ -\frac{1}{2}} = 1\, .
\end{gathered}
\end{equation}
%
A systematic route to a free-field realisation of the chiral $\mathfrak{bms}_4$ algebra is to first realise the larger chiral conformal $\mathfrak{bms}_4$ algebra \cite{Ghorai:2025ebc}, reviewed in  Appendix~\ref{con_bms4}, and then restrict to its chiral $\mathfrak{bms}_4$ subalgebra. The chiral conformal $\mathfrak{bms}_4$ algebra additionally contains a current $D(z)$ with holomorphic weight 1 and two currents $ G_{\pm \frac{1}{2}}^{-}(z)$ with holomorphic weight $\frac{3}{2}$ which are absent from chiral $\mathfrak{bms}_4$. The free-field realisation of chiral conformal $\mathfrak{bms}_4$ presented in Appendix~\ref{con_bms4} is obtained systematically via the BRST approach in terms of three ghost pairs $(\beta_i,\gamma_i)$, and three scalars $\phi_i$, with non-trivial OPEs
 \begin{equation}\label{basic_opes}
\partial\phi_i(z)\,\partial\phi_j(w) \sim -\frac{\delta_{ij}}{(z-w)^2}\,,
\qquad
\gamma_i(z)\,\beta_j(w) \sim \frac{\delta_{ij}}{z-w}\,,
\qquad i,j = 1,2,3\,.
\end{equation}
A free-field realisation of the chiral $\mathfrak{bms}_4$ subalgebra is then easily obtained by restricting to the subset of generators other than $D(z)$ and $ G_{\pm \frac{1}{2}}^{-}(z)$. This realisation will continue to have the same relation between $c$ and $k$ as in the chiral conformal $\mathfrak{bms}_4$. We, however, use a modified realisation of the chiral $\mathfrak{bms}_4$, obtained by dropping/modifying some terms in its realisation as the subalgebra of chiral conformal $\mathfrak{bms}_4$ which do not affect the chiral $\mathfrak{bms}_4$ algebra. In particular, the generators $H^0_a(z)$ and $H^1_\alpha(z) $ of the chiral $\mathfrak{bms}_4$ algebra we work with are
\begin{align}
\label{chir_curr}
H^{0}_{1}(z) &= \beta_1(z)\,,
\nonumber\\[6pt]
H^{0}_{0}(z) &= -\frac{i\sqrt{4+\kappa}}{\sqrt{2}}
\Bigl(\partial\phi_1(z) + \sqrt{3}\,\partial\phi_2(z)\Bigr)
- 2\,(\beta_1\gamma_1)(z) + (\beta_2\gamma_2)(z) - (\beta_3\gamma_3)(z)\,,
\nonumber\\[6pt]
H^{0}_{-1}(z) &= \frac{i\sqrt{4+\kappa}}{\sqrt{2}}\,(\partial\phi_1\,\gamma_1)(z)
+ i\sqrt{\tfrac{3}{2}}\,\sqrt{4+\kappa}\,(\partial\phi_2\,\gamma_1)(z)
+ \bigl(\beta_1(\gamma_1\gamma_1)\bigr)(z)
\nonumber\\
&\quad
- \bigl(\beta_2(\gamma_1\gamma_2)\bigr)(z)
+ \bigl(\beta_3(\gamma_1\gamma_3)\bigr)(z)
+ (\kappa+1)\,\partial\gamma_1(z)\,,
\nonumber\\[6pt]
H^{1}_{\frac{1}{2}}(z) &= -\frac{i\sqrt{4+\kappa}}{\sqrt{2}}\,
\bigl(\partial\phi_1\,\gamma_2\bigr)(z)
+ i\sqrt{\tfrac{3}{2}}\,\sqrt{4+\kappa}\,
\bigl(\partial\phi_2\,\gamma_2\bigr)(z)
- \bigl(\beta_2(\gamma_2\gamma_2)\bigr)(z)
- (\kappa+2)\,\partial\gamma_2(z)\,,
\nonumber\\[6pt]
H^{1}_{-\frac{1}{2}}(z) &= -\bigl\{H^{0}_{-1},\,H^{1}_{\frac{1}{2}}\bigr\}_{1}(z)\,,
\end{align}
where $\{H^{0}_{-1},\, H^{1}_{\frac{1}{2}}\}_{1}$ denotes the residue of the simple pole in the OPE of $H^{0}_{-1}(z)$ with $H^{1}_{\frac{1}{2}}(z)$, and the product $(AB)(z)$ between two fields $A$, $B$  is understood as normal-ordered composite. The holomorphic stress tensor, completing the algebra, is given by
\begin{equation}
\label{freefieldT}
\begin{gathered}
T(z) = -\frac{1}{12(\kappa+4)}\Bigg[6(4+\kappa)
\Bigl((\partial\phi_1\partial\phi_1)(z)
+(\partial\phi_2\partial\phi_2)(z)
+(\partial\phi_3\partial\phi_3)(z)\Bigr)\\
+\sqrt{2}\Bigl(-i\sqrt{4+\kappa}
\bigl(3\kappa\,\partial^2\phi_1(z)
-\sqrt{3}(4+\kappa)\,\partial^2\phi_2(z)\bigr)
+2\sqrt{6}\,\sqrt{\kappa+4}\,(\kappa+7)\,\partial^2\phi_3(z)
\Bigr)\Bigg]\\
+\frac{1}{2}\Bigg[(\partial\beta_2\,\gamma_2)(z)
+(\partial\beta_3\,\gamma_3)(z)
-2\,(\beta_1\partial\gamma_1)(z)
-(\beta_2\partial\gamma_2)(z)
-(\beta_3\partial\gamma_3)(z)\Bigg]\,.
\end{gathered}
\end{equation}
One can verify by direct computation, using \eqref{basic_opes}
that the  generators~\eqref{chir_curr} together with $T(z)$ in (\ref{freefieldT}) satisfy the chiral 
$\mathfrak{bms}_4$ algebra~\eqref{ope_one} with the relation (\ref{candk}). This free-field realisation will serve as the foundation for the construction of graviton primary operators in the following 
section.

Compared to the stress tensor (\ref{stensor}) of the original chiral conformal $\mathfrak{bms}_4$, we have modified the coefficient of the $\partial^2\phi_3$ term in (\ref{freefieldT}). This modification preserves the chiral $\mathfrak{bms}_4$  algebra while changing the expression of the central charge $c$ to that of (\ref{candk}). This modification is necessary for the following reasons. We want to construct a spin-$2$ graviton primary operator $G^{++}_\Delta$  ~\eqref{intro_g_rep}, in terms of the vertex operators of the form $e^{i(r_1\phi_1+r_2\phi_2+r_3\phi_3)}(z)$. 
%
%
The modification we made in the stress tensor makes the conformal dimension of  $G^{++}_\Delta$  linear in the free parameters $r_i$. This linearity will turn out to be essential to obtain the correct OPE structure between two graviton primaries $G^{++}_{\Delta_1}$ and $G^{++}_{\Delta_2}$ as the rhs would involve graviton primaries of the form $G^{++}_{\Delta_1+\Delta_2}$.
\section{Constructing a graviton primary operator}
\label{spin_two_primary}
%
%
We now aim to construct operators in terms of these free-fields ($(\beta_i, \gamma_i)$ and $\phi_i$) that we would like to identify with the graviton operators $G^{++}_\Delta(z, \bar z)$. The guiding principles are $(i)$ they must give the right soft limits 
\begin{eqnarray}
\label{soft-limits}
\lim_{\Delta\to 1} (\Delta -1) G^{++}_\Delta(z, \bar z) 
&=& H^1_{\frac{1}{2}} (z) + \bar{ z} \, H^1_{-\frac{1}{2}}(z)\,, \cr
\lim_{\Delta\to 0} \Delta G^{++}_\Delta(z, \bar z) 
&=& H^0_{1}(z) + \bar {z} \, H^0_0(z)+ \bar{z}^2 \,H^0_{-1}(z)\, .
\end{eqnarray}
and $(ii)$ they must satisfy the expected OPE (\ref{gg_OPE_summary}). 

Our construction of graviton primary operators will be in terms of vertex operators involving the holomorphic free-fields $(\beta_i, \gamma_i)$ and $\phi_i$. In a free-field CFT, vertex 
operators are local operators of the schematic form 
$\mathcal{V}_{h,\bar{h}} = e^{i{\vec r} \cdot {\vec \phi}}$, built from the vector $\vec{\phi}$ of the free scalar fields of 
the theory. A primary vertex operator in our context is a highest-weight Virasoro primary that has a non-singular OPE with the $H^{0}_1(z)$ operator of the current algebra.

Therefore, we begin with the following two highest-weight vertex operators
\begin{equation}
\label{spin_one_vertex}
V_{h,\bar{h}}(z) = \bigl(\beta_1\,{\cal V}\bigr)(z)\,, \qquad
\widetilde{U}_{\tilde{h},\bar{\tilde{h}}}(z) = 
-\bigl\{H^{1}_{-\frac{1}{2}},\,V_{h,\bar{h}}\bigr\}_{1}(z)\,,
\end{equation}
where ${\cal V}\,:=\, e^{i(\sqrt{3}\,r_2\phi_1 + r_2\phi_2 + r_3\phi_3)}$. Note that the structure of $V_{h,\bar{h}}$ is chosen so that in a potential soft limit it reduces to $H^0_1(z)$. For the 
operator $V_{h,\bar{h}}$, the relevant OPEs with the stress tensor and the \(\mathfrak{sl}(2,\mathbb{R})\)
current  $H^{0}_{0}(z)$ are
\begin{equation}
\begin{gathered}
T(z)\,V_{h, \bar h}(w)
\sim \frac{h\,V_{h, \bar h}(w)}{(z-w)^2}
+\frac{\partial_{w}\, V_{h, \bar h}(w)}{z-w},\\[6pt]
H^{0}_{0}(z)\,V_{h, \bar h}(w)
\sim \frac{2\,\bar{h}\,V_{h, \bar h}(w)}{z-w},
\end{gathered}
\end{equation}
which give us its weights $(h, \bar h)$ as 
\begin{align}
h &= 1 + 2r_2^2 + \frac{r_3^2}{2} 
- \frac{(\kappa-2)\,r_2}{\sqrt{6}\,\sqrt{4+\kappa}} 
- \frac{i\,(7+\kappa)\,r_3}{\sqrt{6}\,(4+\kappa)}\,, \nonumber\\
\bar{h} &= -1 - \sqrt{\tfrac{3}{2}}\,\sqrt{4+\kappa}\;r_2\,.
\end{align}
Imposing the helicity condition $h - \bar{h} = 2$ fixes the parameter $r_3$ to be
\begin{equation}
\label{r3_spin1}
r_3 = -2i\,r_2\,,
\end{equation}
leaving $r_2$ as the only free parameter. As advocated for earlier, the scaling dimension $\Delta = h + \bar h$ of $V_{h,\bar{h}}$ 
is now indeed linear in the parameter $r_2$, and is given by
\begin{equation}
\Delta = h + \bar{h} = -\sqrt{6}\,\sqrt{4+\kappa}\;r_2\,.
\end{equation}
For the operator $\widetilde{U}_{\tilde{h},\bar{\tilde{h}}}(z)$, the relevant OPEs 
take the same form as for $V_{h,\bar{h}}$,
\begin{equation}
T(z)\,\widetilde{U}_{\tilde{h},\bar{\tilde{h}}}(w) \sim 
\frac{\tilde{h}\,\widetilde{U}_{\tilde{h},\bar{\tilde{h}}}(w)}{(z-w)^2} 
+ \frac{\partial_{w}\,\widetilde{U}_{\tilde{h},\bar{\tilde{h}}}(w)}{z-w}\,,
\qquad
H^{0}_{0}(z)\,\widetilde{U}_{\tilde{h},\bar{\tilde{h}}}(w) \sim 
\frac{2\,\bar{\tilde{h}}\,\widetilde{U}_{\tilde{h},\bar{\tilde{h}}}(w)}{z-w}\,,
\end{equation}
where
\begin{align}
\tilde{h} &= \frac{3}{2} - \sqrt{\tfrac{3}{2}}\,\sqrt{4+\kappa}\;r_2 
= \frac{\Delta+3}{2}\,, \nonumber\\
\bar{\tilde{h}} &= -\frac{1}{2} - \sqrt{\tfrac{3}{2}}\,\sqrt{4+\kappa}\;r_2 
= \frac{\Delta-1}{2}\,.
\end{align}
One readily verifies that $\tilde{h} - \bar{\tilde{h}} = 2$, confirming that this operator also carries spin $+2$. To bring the two operators onto the same footing (i.e, same $h+\bar h$), 
we perform a shift in $r_2$ and define a new vertex operator
\begin{equation}
U_{h,\bar{h}}(z) = \widetilde{U}_{\tilde{h},\bar{\tilde{h}}}(z)
\Big|_{r_2\,\to\, r_2 + \frac{1}{\sqrt{6}\sqrt{4+\kappa}}}\,,
\end{equation}
whose scaling dimension $h + \bar h$ is also $\Delta$. We thus have constructed two 
highest-weight primary operators $V_{h,\bar{h}}(z)$ and $U_{h,\bar{h}}(z)$, both 
of scaling dimension $\Delta$ and spin/helicity $\sigma = 2$.

 Starting from these two operators, we construct two infinite towers of 
current algebra ($\mathfrak{sl}(2,\mathbb{R})$) descendant (that are still Virasoro primary) vertex operators by acting successively with the lowering operator 
$H^0_{-1}(z)$:
\begin{equation}
    \begin{gathered}
        V_{h,\bar h+1}(z) = -\{H^0_{-1},V_{h,\bar h}\}_1(z); \\ 
        V_{h,\bar h+2}(z)= -\frac{1}{2}\{H^0_{-1},V_{h,\bar h+1}\}_1(z);\\
         V_{h,\bar h+3}(z)= -\mu_{3}\{H^0_{-1},V_{h,\bar h+2}\}_1(z);\\
        \vdots
    \end{gathered}
\end{equation}
and 
\begin{equation}
    \begin{gathered}
        U_{h,\bar h+1}(z) = -\{H^0_{-1},U_{h,\bar h}\}_1(z); \\ 
        U_{h,\bar h+2}(z)= -\nu_{2}\{H^0_{-1},U_{h,\bar h+1}\}_1(z);\\
        U_{h,\bar h+3}(z)= -\nu_{3}\{H^0_{-1},U_{h,\bar h+2}\}_1(z);\\
        \vdots
    \end{gathered}
\end{equation}
The normalisation constants for $\{V_{h,\bar{h}},\,V_{h,\bar{h}+1},\,V_{h,\bar{h}+2}\}$ 
and $\{U_{h,\bar{h}},\,U_{h,\bar{h}+1}\}$ are fixed by requiring that these vertex 
operators reduce to the appropriate currents in the soft limits (\ref{soft-limits}), namely
\begin{equation}
\{V_{h,\bar{h}},\,V_{h,\bar{h}+1},\,V_{h,\bar{h}+2}\}\big|_{\Delta\to 0} 
= \{H^0_{1},\,H^0_0,\,H^0_{-1}\}\,, \qquad
\{U_{h,\bar{h}},\,U_{h,\bar{h}+1}\}\big|_{\Delta\to 1} 
= \{H^1_{\frac{1}{2}},\,H^1_{-\frac{1}{2}}\}\,.
\end{equation}
All other vertex operators have to vanish in these limits. Consequently, the 
remaining normalisation constants $\mu_n$ and $\nu_n$ cannot be fixed by the 
soft limits alone; they will instead be determined later by requiring consistency of the pole structure in the graviton-graviton OPE.

We now assemble the positive-helicity graviton primary operator $G^{++}_{\Delta}(z,\bar{z})$ (i.e, with $\Delta = h + \bar{h}$ and $\sigma = h - \bar{h} = 2$) as the 
following linear combination of vertex operators $U_{h, \bar h}(z)$ and $V_{h, \bar h}(z)$ and their descendants:
\begin{equation}\label{grav_prim}
G^{++}_{\Delta}(z,\bar{z})
= f(\Delta)\,\sum_{n=0}^\infty \bar{z}^n\,U_{h,\bar{h}+n}(z)
\;+\; g(\Delta)\,\sum_{n=0}^\infty \bar{z}^n\,V_{h,\bar{h}+n}(z)\, .
\end{equation}
We adopt the notation 
$G^{++}_\Delta(z,\bar{z})$ in place of $G_{h,\bar{h}}(z)$, as the former is more 
standard in the celestial holography literature. The normalisation functions $f(\Delta)$ 
and $g(\Delta)$ are constrained by demanding that the graviton primary reproduces the 
correct soft currents in the leading and subleading soft limits \eqref{soft-limits}.
These two conditions translate into the following constraints on $f(\Delta)$ and 
$g(\Delta)$:
\begin{equation}
    \begin{split}
  \lim_{\Delta\to 1}\,   (\Delta-1)\, f(\Delta) &= 1, \quad  \lim_{\Delta\to 0}\,   \Delta\, f(\Delta) = 0,\\[6pt]
   \lim_{\Delta\to 1}\,   (\Delta-1)\, g(\Delta) &= 0, \quad  \lim_{\Delta\to 0}\,   \Delta\, g(\Delta) = 1\,.\\
    \end{split}
\end{equation}
In addition to the graviton primary, we introduce a highest-weight scalar primary operator, which will play an important role in the graviton-graviton OPE:
\begin{equation}\label{scalar_primary}
\Phi_{\Delta}(z,\bar{z}) =\, s(\Delta)\,\sum_{n=0}^{\infty}\bar{z}^n\,
{\cal V}_{h,\bar{h}+n}(z)\,,
\end{equation}
where the component operators are defined recursively by
\begin{equation*}
\begin{gathered}
{\cal V}_{h,\bar{h}}(z)=e^{i \, r_2 \, (\sqrt{3} \, \phi_1+\phi_2-2i\,\phi_3)}(z)\,,\\
{\cal V}_{h,\bar{h}+1}(z)=-\{H^0_{-1},{\cal V}_{h,\bar{h}}\}_1(z)\,,\\
\vdots
\end{gathered}
\end{equation*}
From the soft theorem analysis of bulk amplitudes in the BW 
theory\footnote{Following the notation of~\cite{Cachazo:2014fwa}, consider the 
$(n+1)^{\text{th}}$ particle to be a scalar $\phi^{-+}_z$ with $h = 0$ in the BW 
theory~\cite{Ghorai:2025ebc}. In this case,
\begin{align*}
\mathcal{M}^{12}_{n+1}\!\left(\{\lambda_1,\cdots,\lambda_n,\epsilon\lambda_s\},
\{\tilde{\lambda}_3,\cdots,\tilde{\lambda}_n,\tilde{\lambda}_s\}\right)
= -\mathcal{M}^{(12)}_n\!\left(\{\lambda_1,\cdots,\lambda_n\},
\{\tilde{\lambda}_3,\cdots,\tilde{\lambda}_n\}\right) + \mathcal{O}(\epsilon)\,.
\end{align*}} at subleading order, the scalar primary reduces in the soft limit, to  
\begin{equation}\label{soft_limit_scalar}
\lim_{\Delta\to 0}\Delta\,\Phi_{\Delta}(z,\bar{z}) = -\mathbb{I}\,.
\end{equation}
where $\mathbb{I}$ is the identity 
operator. This condition fixes the behaviour of the normalisation function $s(\Delta)$ near 
$\Delta = 0$ to be 
\begin{equation}
\lim_{\Delta\to 0}\Delta\,s(\Delta) = -1\,.
\end{equation}
So far we have used the constraints coming from the soft-limits. Next, we will utilise conditions coming from OPEs.
\section{\texorpdfstring{$G^{++}_{\Delta_1}(z,\bar{z})\,G^{++}_{\Delta_2}(w,\bar{w})$}{gg} OPE}\label{ope_two_graviton}
%
Having constructed the graviton primary operator in terms of vertex operators, we now 
turn to the computation of the OPE between two positive-helicity graviton primaries. 
Expanding the product using the definition~\eqref{grav_prim}, we obtain
\begin{align}\label{gg_expansion}
G^{++}_{\Delta_1}(z,\bar{z})\,G^{++}_{\Delta_2}(w,\bar{w})
&= f(\Delta_1)\,f(\Delta_2)\sum_{n,m=0}^\infty \bar{z}^n\bar{w}^m\,
U_{h_1,\bar{h}_1+n}(z)\,U_{h_2,\bar{h}_2+m}(w)
\notag\\
&+\,f(\Delta_1)\,g(\Delta_2)\sum_{n,m=0}^\infty \bar{z}^n\bar{w}^m\,
U_{h_1,\bar{h}_1+n}(z)\,V_{h_2,\bar{h}_2+m}(w)
\notag\\
&+\,g(\Delta_1)\,f(\Delta_2)\sum_{n,m=0}^\infty \bar{z}^n\bar{w}^m\,
V_{h_1,\bar{h}_1+n}(z)\,U_{h_2,\bar{h}_2+m}(w)
\notag\\
&+\,g(\Delta_1)\,g(\Delta_2)\sum_{n,m=0}^\infty \bar{z}^n\bar{w}^m\,
V_{h_1,\bar{h}_1+n}(z)\,V_{h_2,\bar{h}_2+m}(w)\,.
\end{align}
Our goal is to first write the expansion in $(\bar z, \bar w)$ explicitly in the above equation and then at each order take the holomorphic OPE limit  $z \to w$. After that, we take the OPE limit in the anti-holomorphic variable $\bar z \to \bar w$ as well, and verify that it reproduces the 
graviton-graviton OPE obtained from the collinear limit of the Mellin-transformed MHV 
amplitudes in conformal gravity, as given in \eqref{gg_OPE_summary}.\footnote{However, while taking the anti-holomorphic OPE limit $\bar z \to \bar w$, we must be careful not to drop any positive powers of $(\bar z - \bar w)$ that are still singular at vanishing $(z-w)$.}

Let us proceed by analysing each line in \eqref{gg_expansion}.  A somewhat surprising feature of the free-field realisation is that the OPE between any two $U_{h,\bar h}$ vertex operators is non-singular,
\begin{equation}
U_{h_1,\bar{h}_1+m}(z)\,U_{h_2,\bar{h}_2+n}(w) \sim 0\,, 
\qquad n,m \in \mathbb{Z}_{\geq 0}\,.
\end{equation}
As a consequence, the first line of \eqref{gg_expansion} does not contribute to 
the singular part of the OPE, and the entire non-trivial content of the 
graviton-graviton OPE is carried by the mixed $U$-$V$ and $V$-$V$ sectors.

 We will write down the OPE expansion for the mixed sector \(U\)-\(V\) and \(V\)-\(V\) order by order in $\left(\bar z, \bar w \right)$. 
 So, first we start with the order zero term, i.e. \(\mathcal{O}( \bar w^0)\) term.
 
\subsection{\texorpdfstring{$\mathcal{O}( \bar w^0)$}{oz0} terms}
\begin{equation}
\begin{gathered}
        G^{++}_{\Delta_1}(z,\bar z) G^{++}_{\Delta_2}(w,\bar w)\Bigg|_{\mathcal{O}(\bar w^0)}  = f(\Delta_1)  g(\Delta_2) U_{h_{1}, \bar {h}_{1}}(z)  V_{h_2, \bar h_2}(w) + g(\Delta_1)  f(\Delta_2) V_{h_1, \bar h_1}(z)  U_{h_2, \bar h_2}(w)\\
        + g(\Delta_1) g(\Delta_2) V_{h_1, \bar h_1}(z)  V_{h_2, \bar h_2}(w)\,.
\end{gathered}
\end{equation}
One can also check that
\begin{equation}
    \begin{gathered}
        U_{h_1, \bar h_1}(z)  V_{h_2, \bar h_2}(w) \sim 0, \ V_{h_1, \bar h_1}(z)  U_{h_2, \bar h_2}(w) \sim 0, \ V_{h_1, \bar h_1}(z)  V_{h_2, \bar h_2}(w) \sim 0.
    \end{gathered}
\end{equation}
Hence,
\begin{equation}
        G^{++}_{\Delta_1}(z,\bar z) G^{++}_{\Delta_2}(w,\bar w)\Bigg|_{\mathcal{O}(\bar w^0)}  \sim 0.
\end{equation}

\noindent Now the first non-trivial contribution comes from the order \(\mathcal{O}(\bar w)\) term.
\subsection{\texorpdfstring{$\mathcal{O}(\bar w)$}{Oz} terms}
\begin{equation}
\begin{gathered}
    G^{++}_{\Delta_1}(z,\bar z) G^{++}_{\Delta_2}(w,\bar w)\Bigg|_{\mathcal{O}(\bar w)} =\, f(\Delta_1)  g(\Delta_2)  \left[ \bar z \, U_{h_1, \bar h_1+1}(z)  V_{h_2, \bar h_2}(w) + \bar w \, U_{h_1, \bar h_1}(z)  V_{h_2, \bar h_2+1}(w) \right] \\[6pt]
    +\,\, g(\Delta_1)  f(\Delta_2) \left[  \bar z \, V_{h_1, \bar h_1+1}(z)  U_{h_2, \bar h_2}(w) + \bar w \, V_{h_1, \bar h_1}(z)  U_{h_2, \bar h_2+1}(w) \right] \\[8pt]
    +\,\, g(\Delta_1) g(\Delta_2) \left[  \bar z \, V_{h_1, \bar h_1+1}(z)  V_{h_2, \bar h_2}(w) + \bar w \, V_{h_1, \bar h_1}(z)  V_{h_2, \bar h_2+1}(w) \right].
    \end{gathered}
\end{equation}
The OPEs between the $U$ and $V$-vertex operators appearing in the above equation are given by
\begin{equation*}
\begin{gathered}
   U_{h_1, \bar h_1+1}(z)  V_{h_2, \bar h_2}(w) \sim (\Delta_1-2)\frac{U_{h_1+h_2-1, \bar h_1+\bar h_2 + 1}(w)}{z-w}\,,\\[4pt]
   U_{h_1, \bar h_1}(z)  V_{h_2, \bar h_2+1}(w) \sim -(\Delta_1-2)\frac{U_{h_1+h_2-1, \bar h_1+\bar h_2+1}(w)}{z-w}\,,\\[4pt]
    V_{h_1, \bar h_1+1}(z)  U_{h_2, \bar h_2}(w) \sim (\Delta_2-2)\frac{U_{h_1+h_2-1, \bar h_1+\bar h_2 + 1}(w)}{z-w}\,,\\[4pt]
    V_{h_1, \bar h_1}(z)  U_{h_2, \bar h_2+1}(w) \sim -(\Delta_2-2)\frac{U_{h_1+h_2-1, \bar h_1+\bar h_2 + 1}(w)}{z-w}\,,\\[4pt]
    V_{h_1, \bar h_1+1}(z)  V_{h_2, \bar h_2}(w) \sim (\Delta_1+\Delta_2-2)\frac{V_{h_1+h_2-1,\bar h_1+\bar h_2+1}(w)}{z-w}\,,\\[4pt]
    V_{h_1, \bar h_1}(z)  V_{h_2, \bar h_2+1}(w) \sim -(\Delta_1+\Delta_2-2)\frac{V_{h_1+h_2-1,\bar h_1+\bar h_2+1}(w)}{z-w}.
    \end{gathered}
\end{equation*}
Using these OPEs we can compute the $\mathcal{O}(\bar w)$ term in graviton-graviton OPE:
\begin{equation}
\begin{gathered}
    G^{++}_{\Delta_1}(z,\bar z) G^{++}_{\Delta_2}(w,\bar w)\Bigg|_{\mathcal{O}(\bar w)}    = \frac{(\bar z - \bar w)}{(z-w)} \Bigg[ E_U(\Delta_1,\Delta_2)\, f(\Delta_1+\Delta_2)\, U_{h_1+h_2-1, \bar h_1+\bar h_2 + 1}(w) \\[4pt]
    \,+\,  E_V(\Delta_1,\Delta_2)\, g(\Delta_1+\Delta_2)\, V_{h_1+h_2-1, \bar h_1+\bar h_2 + 1}(w) \Bigg]\,,
    \end{gathered}
\end{equation}
where 
\begin{equation}
\begin{gathered}
    E_U(\Delta_1,\Delta_2) = \frac{f(\Delta_1)  g(\Delta_2)(\Delta_1-2)   + g(\Delta_1)  f(\Delta_2)(\Delta_2-2)}{f(\Delta_1+\Delta_2)}\,,\\[6pt]
     E_V(\Delta_1,\Delta_2) = \frac{g(\Delta_1) g(\Delta_2) (\Delta_1+\Delta_2-2)}{g(\Delta_1+\Delta_2)}\,.
     \end{gathered}
\end{equation}

\noindent Proceeding to the next order, we analyse the $\mathcal{O}(\bar w^2)$ 
contribution to the graviton-graviton OPE.
\subsection{\texorpdfstring{$\mathcal{O}(\bar w^2)$}{Oz2} terms}
\begin{equation}\label{order_O_2}
\begin{gathered}
    G^{++}_{\Delta_1}(z,\bar{z})\,G^{++}_{\Delta_2}(w,\bar{w})\bigg|_{\mathcal{O}(\bar w^2)}= \sum_{n=0}^{2}\bar{z}^{\,n}\bar{w}^{\,2-n}\Bigg[
f(\Delta_1)\,g(\Delta_2)\,U_{h_1,\bar{h}_1+n}(z)\,V_{h_2,\bar{h}_2+2-n}(w)\notag\\
+\,g(\Delta_1)\,f(\Delta_2)\,V_{h_1,\bar{h}_1+n}(z)\,U_{h_2,\bar{h}_2+2-n}(w)
+\,g(\Delta_1)\,g(\Delta_2)\,V_{h_1,\bar{h}_1+n}(z)\,V_{h_2,\bar{h}_2+2-n}(w)
\Bigg]\,.
\end{gathered}
\end{equation}
To evaluate this, we need the OPEs between all the $U$-$V$ and $V$-$V$ vertex operators appearing in the above equation.  Two representative examples of the $U$-$V$ OPEs are
\begin{align}
U_{h_1,\bar{h}_1+2}(z)\,V_{h_2,\bar{h}_2}(w)
&\sim \phantom{-}\frac{2\nu_2\,(\Delta_1-2)\,(\Delta_1-1)}
{\Delta_1+\Delta_2-1}
\frac{U_{h_1+h_2-1,\bar{h}_1+\bar{h}_2+2}(w)}{z-w}\,,
\\[4pt]
U_{h_1,\bar{h}_1+1}(z)\,V_{h_2,\bar{h}_2+1}(w)
&\sim -\frac{(\Delta_1-2)\,(\Delta_1-\Delta_2)}
{\Delta_1+\Delta_2-2}
\frac{U_{h_1+h_2-1,\bar{h}_1+\bar{h}_2+2}(w)}{z-w}\,,
\end{align}
with the remaining $U$-$V$ OPEs collected in Appendix~\ref{order_o2}. Among the 
$V$-$V$ OPEs, we will also write one of the representatives
\begin{align}
V_{h_1,\bar{h}_1+2}(z)\,V_{h_2,\bar{h}_2}(w)
&= (\Delta_1-1)\frac{V_{h_1+h_2-1,\bar{h}_1+\bar{h}_2+2}(w)}{z-w}
+\alpha\,\frac{({\cal V}_1{\cal V}_2)(w)}{(z-w)^2}
+\cdots\,,
\end{align}
where $\alpha = -1 + \Delta_1(\Delta_2-1) - \Delta_2 - \kappa$, and the ellipses denote descendant contributions. We have also defined\footnote{For the rest of this paper, we use the shorthand ${\cal V}_i(z) \equiv {\cal V}_{h_i,\bar{h}_i}(z)$, where the subscript $i$ labels the conformal weights $(h_i, \bar{h}_i)$ associated with the scaling dimension $\Delta_i$.}
\begin{equation*}
({\cal V}_1{\cal V}_2)(z)=e^{i \, (r_2+s_2) \, (\sqrt{3} \, \phi_1+\phi_2-2i\,\phi_3)}(z) \, .
\end{equation*} 
The remaining $V$-$V$ OPEs are again relegated to Appendix~\ref{order_o2}. 
A notable feature of this order 
is the appearance of a double-pole singularity, whose coefficient is precisely the 
scalar primary $\Phi_\Delta$ introduced in Section~\ref{spin_two_primary}; a 
distinctive feature of conformal gravity with no counterpart in Einstein gravity.

\noindent Using these OPEs and setting $\nu_2 = \tfrac{1}{2}$, the $\mathcal{O}(\bar w^2)$ contribution takes the form
{\small 
\begin{equation}
\begin{gathered}
 G^{++}_{\Delta_1}(z,\bar z) G^{++}_{\Delta_2}(w,\bar w)\Bigg|_{\mathcal{O}( \bar w^2)} =\frac{(\bar z-\bar w)}{(z-w)}\Big[E_U(\Delta_1, \Delta_2)\, f(\Delta_1+\Delta_2)\, \bar{w}\,U_{h_1+h_2-1,\bar {h}_1+\bar{h}_2+2}(w)\\[-4pt]
 +E_V(\Delta_1, \Delta_2)\, g(\Delta_1+\Delta_2)\,\bar{w}\, V_{h_1+h_2-1,\bar {h}_1+\bar{h}_2+2}\Big]
 +\left(\frac{\bar z-\bar w}{z-w}\right)^2 \mathcal{Q}(\Delta_1, \Delta_2) \, s(\Delta_1+\Delta_2)\, ({\cal V}_1{\cal V}_2)(w)\,,
\end{gathered}
\end{equation}
}
 where
\begin{equation}
\begin{gathered}
    \mathcal{Q}(\Delta_1, \Delta_2)=\frac{\left(\Delta_1(\Delta_2-1)-\Delta_2-1-\kappa\right)g(\Delta_1)g(\Delta_2)}{s(\Delta_1+\Delta_2)}.
\end{gathered}
\end{equation}
%
\subsection{\texorpdfstring{$\mathcal{O}( \bar w^3)$}{Oz3} terms}
%
%
\begin{equation}\label{order_O_3}
\begin{gathered}
    G^{++}_{\Delta_1}(z,\bar{z})\,G^{++}_{\Delta_2}(w,\bar{w})\bigg|_{\mathcal{O}(\bar w^3)}= \sum_{n=0}^{3}\bar{z}^{\,n}\bar{w}^{\,3-n}\Bigg[
f(\Delta_1)\,g(\Delta_2)\,U_{h_1,\bar{h}_1+n}(z)\,V_{h_2,\bar{h}_2+3-n}(w)\\[-4pt]
+\,g(\Delta_1)\,f(\Delta_2)\,V_{h_1,\bar{h}_1+n}(z)\,U_{h_2,\bar{h}_2+3-n}(w)
+\,g(\Delta_1)\,g(\Delta_2)\,V_{h_1,\bar{h}_1+n}(z)\,V_{h_2,\bar{h}_2+3-n}(w)
\Bigg]\,.
\end{gathered}
\end{equation}
\noindent Proceeding in a similar manner as before, we will compute the OPE expansion for each term in the above ~\eqref{order_O_3}. The details of the computation are given in the appendix~\ref{order_o3}. Using the results of appendix~\ref{order_o3}, and setting $\mu_3 = \tfrac{1}{3}$ and $\nu_3 = \tfrac{1}{3}$, we obtain the $\mathcal{O}( \bar w^3)$ contribution to the graviton-graviton OPE as,
\begin{eqnarray}
 && G^{++}_{\Delta_1}(z,\bar z) G^{++}_{\Delta_2}(w,\bar w)\Bigg|_{\mathcal{O}(\bar w^3)} \!\!\!\!\!\!\!\! =\left(\frac{\bar z-\bar w}{z-w}\right)\Big[E_U(\Delta_1,\Delta_2)\,\, f(\Delta_1+\Delta_2)\,\bar{w}^2\,U_{h_1+h_2-1,\bar {h}_1+\bar{h}_2+3}(w)\cr
&&  ~~~~~~~~~~~~ +\,E_V(\Delta_1,\Delta_2)\,g(\Delta_1+\Delta_2)\,\bar{w}^2\, V_{h_1+h_2-1,\bar {h}_1+\bar{h}_2+3}\Big]\cr
&&  ~~~~~~~~~~~~ +\,\left(\frac{\bar z-\bar w}{z-w}\right)^2 \mathcal{Q}(\Delta_1, \Delta_2)\, s(\Delta_1+\Delta_2)\,(\Delta_1+\Delta_2)\,\bar{w} \, (\gamma_1{\cal V}_1{\cal V}_2)(w).
\end{eqnarray}
\noindent Collecting all contributions up to $\mathcal{O}(\bar{w}^3)$, the 
graviton-graviton OPE takes the form
\begin{equation}\label{graviton_graviton_ope_order_three}
\begin{gathered}
 G^{++}_{\Delta_1}(z,\bar z) G^{++}_{\Delta_2}(w,\bar w)
 =\left(\frac{\bar z-\bar w}{z-w}\right)\Bigg[E_U(\Delta_1,\Delta_2)\,\, f(\Delta_1+\Delta_2)\,\Big(U_{h_1+h_2-1,\bar {h}_1+\bar{h}_2+1}(w)\\[-4pt]
 ~~~~ +\bar{w}\,U_{h_1+h_2-1,\bar {h}_1+\bar{h}_2+2}(w)+ \bar{w}^2\,U_{h_1+h_2-1,\bar {h}_1+\bar{h}_2+3}(w)\Big)\\[-4pt]
 +E_V(\Delta_1,\Delta_2)\,\, g(\Delta_1+\Delta_2)\,\Big(V_{h_1+h_2-1,\bar {h}_1+\bar{h}_2+1} (w) \\[-4pt]
~~~~ +\bar{w}\,V_{h_1+h_2-1,\bar {h}_1+\bar{h}_2+2}(w)+\bar{w}^2\, V_{h_1+h_2-1,\bar {h}_1+\bar{h}_2+3}(w)\Big)\Bigg]\\[-4pt]
~~~~ +\left(\frac{\bar z-\bar w}{z-w}\right)^2 \mathcal{Q}(\Delta_1,\Delta_2)\,s(\Delta_1+\Delta_2)\,\Big[({\cal V}_1{\cal V}_2)(w)+\bar{w}\,(\Delta_1+\Delta_2)\, (\gamma_1{\cal V}_1{\cal V}_2)(w)\Big] + \mathcal{O}(\bar{w}^4).
\end{gathered}
\end{equation}
 \noindent The right-hand side of the above equation describes how two gravitons fuse into a series, which can produce a graviton primary of dimension 
$\Delta_1+\Delta_2$ with appropriate choice of the normalisation factors, and the presence of a series for the scalar primary $\Phi_\Delta$ in the OPE channel. As we will see shortly, once the normalisation functions are fixed, this order-by-order structure resums beautifully into a simple closed-form expression given in \eqref{gg_OPE_summary}.

We now determine the normalisation functions $f(\Delta)$, $g(\Delta)$ and $s(\Delta)$ introduced 
in \eqref{grav_prim} and~\eqref{scalar_primary}. A natural and self-consistent 
choice, is to express them in terms of Euler gamma functions:
\begin{equation}
\begin{gathered}
f(\Delta)=-\Delta\,\Gamma(\Delta-2)\,,\quad
g(\Delta)=\frac{\kappa+1}{\Delta+\kappa+1}\,\Gamma(\Delta)\,,
\end{gathered}
\end{equation}
and
\begin{equation}
s(\Delta)=-\Gamma(\Delta)\,.
\end{equation}
%
%
With these normalisations in hand, the graviton-graviton OPE in 
\eqref{graviton_graviton_ope_order_three}, truncated at $\mathcal{O}(\bar{w}^3)$, 
takes the form
\begin{equation}
\begin{gathered}
G^{++}_{\Delta_1}(z,\bar z) G^{++}_{\Delta_2}(w,\bar w) 
=-\left(\frac{\bar z-\bar w}{z-w}\right)\,B(\Delta_1-1,\Delta_2-1)\,\Bigg[\\[6pt]
f(\Delta_1+\Delta_2)\,\Big(
U_{h_1+h_2-1,\bar {h}_1+\bar{h}_2+1}(w)
+\bar{w}\,U_{h_1+h_2-1,\bar {h}_1+\bar{h}_2+2}(w)
+\bar{w}^2\,U_{h_1+h_2-1,\bar {h}_1+\bar{h}_2+3}(w)
\Big)\\[-4pt]
+\, g(\Delta_1+\Delta_2)\,\Big(
V_{h_1+h_2-1,\bar {h}_1+\bar{h}_2+1}(w)
+\bar{w}\,V_{h_1+h_2-1,\bar {h}_1+\bar{h}_2+2}(w)
+\bar{w}^2\,V_{h_1+h_2-1,\bar {h}_1+\bar{h}_2+3}(w)
\Big)
\Bigg]\\
+\left(\frac{\bar z-\bar w}{z-w}\right)^2\, 
B(\Delta_1,\Delta_2)\, s(\Delta_1+\Delta_2)\,\Big[
({\cal V}_1{\cal V}_2)(w)
+\bar{w}\,(\Delta_1+\Delta_2)\, (\gamma_1{\cal V}_1{\cal V}_2)(w)
\Big] + \mathcal{O}(\bar {w}^4).
\end{gathered}
\end{equation}
In the final step above, we have taken $\kappa \to -2$, only in which the Euler beta function emerges in the graviton-graviton OPE expansion. This analysis is expected to hold at arbitrarily high orders in $\bar{w}$. 

Resumming the full series and using the representations of the graviton and scalar primaries 
given in \eqref{grav_prim} and~\eqref{scalar_primary}, together with the  identification
\begin{equation*}
(\Delta_1+\Delta_2)\,{(\gamma_1{\cal V}_1{\cal V}_2)}(z) 
= -\bigl\{H^0_{-1},\,{({\cal V}_1{\cal V}_2)}\bigr\}_1(z)\,,
\end{equation*}
we obtain the closed-form graviton-graviton OPE:
\begin{align}\label{com_graviton_graviton_ope}
G^{++}_{\Delta_1}(z,\bar{z})\,G^{++}_{\Delta_2}(w,\bar{w})
&= -\frac{(\bar{z}-\bar{w})}{(z-w)}
B(\Delta_1-1,\Delta_2-1)\,G^{++}_{\Delta_1+\Delta_2}(w,\bar{w})
\notag\\[4pt]
&\qquad
-\frac{(\bar{z}-\bar{w})^2}{(z-w)^2}\,
B(\Delta_1,\Delta_2)\,\Phi_{\Delta_1+\Delta_2}(w,\bar{w})
+\cdots\,,
\end{align}
 \noindent where $\cdots$ represents the descendants. This result is in exact agreement with the OPE \eqref{gg_OPE_summary} derived directly from bulk MHV amplitudes in the BW  theory~\cite{Ghorai:2025ebc} with the first term accounting for the exchange of a graviton primary, 
while the second term reflects the contribution of the scalar primary $\Phi_\Delta$,  whose appearance is expected of BW conformal gravity. This completes our construction of the graviton vertex operator in terms of free fields $(\beta_i, \gamma_i)$ and $\phi_i$.
%
\subsubsection*{OPE between the graviton primary and an outgoing scalar primary}
\label{graviton_scalar_ope}
As a powerful check of our definitions of the positive helicity graviton operator and the scalar operator, we compute the OPE between the graviton primary 
$G^{++}_{\Delta_1}$ and the scalar primary $\Phi_{\Delta_2}$. As we have done in the section~\ref{spin_two_primary}, we expand the scalar primary operator as
\begin{equation*}
\Phi_{\Delta}(z,\bar{z}) = -\Gamma(\Delta) \Bigl(
{\cal V}_{h,\bar h}(z)
+\bar{z}\,{\cal V}_{h,\bar{h}+1}(z)
+\bar{z}^2\,{\cal V}_{h,\bar{h}+2}(z)
+\cdots\Bigr)\,.
\end{equation*}
Using this expansion, and noting that the first part of the graviton operator, \(\sum_{n=0}^\infty \bar{z}^n\, U_{h,\bar{h}+n}(z)\) commutes with 
$\Phi_{\Delta}(z,\bar{z})$, the relevant components of positive helicity graviton-scalar OPE are
\begin{equation}
\begin{gathered}
    V_{h_1,\bar {h}_1}(z,\bar{z})\,{\cal V}_{h_2,\bar{h}_2+1}(w,\bar{w})
=
-\Delta_2\,\frac{{\cal V}_{h+h_2-1,\bar{h}_1+\bar{h}_2+1}(w,\bar{w})}{z-w}\,,\\[6pt]
V_{h_1,\bar {h}_1+1}(z,\bar{z})\,{\cal V}_{h_2,\bar{h}_2}(w,\bar{w})
=
\Delta_2\,\frac{{\cal V}_{h+h_2-1,\bar{h}_1+\bar{h}_2+1}(w,\bar{w})}{z-w}\,, ~~\text{etc.}
\end{gathered}
\end{equation}
Combining these contributions, and again taking the limit $\kappa \to -2$, we obtain
\begin{equation}
\label{com_graviton_scalar_2}
G^{++}_{\Delta_1}(z,\bar{z})\,\Phi_{\Delta_2}(w,\bar{w})
= -\frac{(\bar{z}-\bar{w})}{(z-w)}\,
B(\Delta_1-1,\Delta_2+1)\,\Phi_{\Delta_1+\Delta_2}(w,\bar{w}) + \cdots\,.
\end{equation}
This expression is in exact agreement with (\ref{com_graviton_scalar}), 
providing a further non-trivial consistency check of our celestial CFT construction. Finally, it is evident that the OPE between two scalar primaries $\Phi_\Delta$ is non-singular, again as expected for the BW theory. 

This completes the demonstration that the graviton and scalar primaries we have constructed do reproduce the expected OPEs and hence are good celestial holographic duals of the corresponding bulk particles of the BW theory.
%
%
\section{Discussion}\label{discussion}
%
In this paper, we have constructed a candidate celestial CFT dual for the MHV sector of conformal gravity (the bosonic subsector of the Berkovits--Witten string 
theory). Using the free-field representation of the chiral $\mathfrak{bms}_4$ symmetry algebra, we identified primary vertex operators of this algebra and using them, we have explicitly constructed the positive-helicity graviton primary $G^{++}_{\Delta}$ and a scalar primary $\Phi_{\Delta}$, both transforming covariantly under the symmetry. The resulting graviton-graviton, graviton-scalar OPEs are in exact agreement with the Mellin transform of the collinear limit of 
 MHV conformal gravity bulk amplitudes~\cite{Ghorai:2025ebc}, and the $\mathfrak{sl}(2,\mathbb{R})$ current algebra fixed at level $\kappa = -2$. This provides a concrete and self-consistent celestial dual description of the MHV 
sector of conformal gravity. 

 Furthermore, the chiral $\mathfrak{bms}_4$ algebra has a non-vanishing central extension, as shown in \eqref{central_term_current_algebra}. 
This central term is a novel feature that distinguishes conformal gravity from  Einstein gravity~\cite{Banerjee:2020zlg}, and can be traced directly to the fact that in the soft limit 
$\Delta \to 0$, the scalar primary $\Phi_\Delta$ appearing in the graviton-graviton 
OPE reduces to the (negative of the) identity operator ($-\mathbb{I}$), as shown in \eqref{soft_limit_scalar}.

The results presented in this work open up several natural and important avenues for future investigation. An immediate direction is the computation of full correlation functions in the dual celestial CFT. The OPE derived in this work captures the structure of scattering amplitudes only in the collinear limit. A complete match between bulk scattering amplitudes and boundary observables requires going beyond this limit and computing full $n$-point correlation functions on the celestial sphere. 

Another direction is the systematic computation of the full set of OPEs among all the operators of the theory. So far only $G^{++} G^{++}$ and $G^{++}\Phi$ and $\Phi \Phi$ OPEs have been computed. The full story requires knowing all combinations $G^{++}G^{--},  G^{--}\Phi$, and $G^{--} G^{--}$. This will involve the construction of the negative helicity graviton operator, which we still lack.


An interesting question is the construction of an explicit two-dimensional action for the celestial CFT. The free field realisation employed here provides a concrete representation of the symmetry algebra and has proven powerful enough to reproduce the graviton OPE of conformal gravity, but it does not by itself constitute a Lagrangian description of the theory. A gauged WZW model (which depends on the embedding of $\mathfrak{sl}(2,\mathbb{R})$ and the corresponding constraints to be imposed) based on the gauge group $\mathfrak{so}(2,4)$ (or $\mathfrak{sl}(4,\mathbb{R})$) is a natural candidate for such an action that would admit the chiral conformal $\mathfrak{bms}_4$ symmetry algebra (which is the ${\cal W}$-algebra obtained by an appropriate DS reduction of the $\mathfrak{sl}(4,\mathbb{R})$ current algebra), which in turn means it would admit the chiral $\mathfrak{bms}_4$. Such a gauge WZW model can indeed be written down on the lines of Bershadsky \cite{Bershadsky:1990bg} and Bershadsky-Ooguri \cite{Bershadsky:1989mf}. However, we have modified the expression of the charges of the chiral $\mathfrak{bms}_4$ subalgebra to suit our purposes as explained in section \ref{free_field_rep}, and it would be interesting to seek a Lagrangian that admits these modified charges as its conserved charges. 

The bulk amplitudes from which we extracted the OPE of graviton operators were at tree level. It is natural to ask what, if anything, will change once loops in the bulk are considered for our investigations from the CFT perspective. In this connection, we recall that the bulk answers (such as (\ref{gg_OPE_summary}) and (\ref{com_graviton_scalar})) matched with the boundary ones under the $\kappa \rightarrow -2$ limit. What about terms involving powers of $\kappa +2$? Do they correspond to going beyond tree-level on the bulk side? We want to pursue this direction in the future. 

Some other questions that arise from this work include the following: $(i)$
The bulk BW theory is superconformal. Where is supersymmetry/conformal symmetry on the boundary side? $(ii)$ The tree-level MHV sector of Einstein gravity has been shown to admit a much bigger symmetry algebra than the chiral $\mathfrak{bms}_4$, namely the  $w_{1+\infty}$ by Strominger et al. Is there such an extension in the current context?. We will leave such issues for the future. 
%
\section*{Acknowledgements} 
%
We would like to thank the participants of the Chennai String Meeting (held during March 2026 at IMSc, Chennai), for their insightful questions and discussions, during which preliminary results of this work were presented. We thank Dileep Jatkar and Alok Laddha for helpful discussions during the course of this work.
\appendix
%
\section{Free-field theory of chiral conformal \texorpdfstring{\(\mathfrak{bms}_4\)}{bms4} algebra}
\label{con_bms4}
%
In this section, we present the free-field realisation of the chiral conformal $\mathfrak{bms}_4$ algebra. Its operator content consists of a chiral $\mathfrak{sl}(2,\mathbb{R})$ current algebra generated by currents $H^0_a(z)$ with $a = 0, \pm 1$, a  chiral current $D(z)$ with holomorphic weight 1, four chiral primary operators $H^1_{\pm \frac{1}{2}}(z)$ and $G^-_{\pm \frac{1}{2}}(z)$ with holomorphic weight $\frac{3}{2}$, and a chiral stress tensor $T(z)$. 
The OPEs among the chiral operators introduced above are given by
\begin{equation}
\begin{gathered}
\nonumber T(z)T(w) \sim \frac{c/2}{(z-w)^4}+\frac{2 T(w)}{(z-w)^2}+\frac{\partial_w T(w)}{z-w},\\
     H^0_a(z)H^0_b(w) \sim \frac{-\frac{k}{2} \eta_{ab}}{(z-w)^2}+\frac{f^{c}_{~ab}~H^0_c(w)}{z-w},\\
     D(z)D(w) \sim \frac{1}{(z-w)^2}, \quad
    D(z)H^0_a(w) \sim 0,\\
      H^0_a(z)H_{\alpha}^{1}(w) \sim \frac{(\lambda_a)^\beta_{~~\alpha}H_\beta^{1}(w)}{z-w},\quad
    H^0_a(z)G_\alpha^{-}(w)\sim \frac{(\lambda_a)^{\beta}_{~~\alpha}G_\beta^{-}(w)}{z-w},\\
     D(z)H_\alpha^{1}(w)\sim \frac{q~H_\alpha^{1}(w)}{z-w}, \qquad   D(z)G_\alpha^{-}(w)\sim -\frac{q~G_\alpha^{-}(w)}{z-w},
    \end{gathered}
\end{equation}
\begin{equation}
\label{ope_ones}
\begin{gathered} 
    H_\alpha^{1}(z)H_{\beta}^{1}(w)\sim 0, \quad
    G_{\alpha}^{-}(z)G_{\beta}^{-}(w)\sim 0,\\
    T(z)H_\alpha^{1}(w) \sim \frac{\tfrac{3}{2} H_\alpha^{1}(w)}{(z-w)^2}+\frac{\partial_w H_\alpha^{1}(w)}{z-w}, \qquad T(z)G_\alpha^{-}(w) \sim \frac{\tfrac{3}{2} G_\alpha^{-}(w)}{(z-w)^2}+\frac{\partial_w G_\alpha^{-}(w)}{z-w}\\
     T(z)H^0_a(w) \sim \frac{H^0_a(w)}{(z-w)^2}+\frac{\partial_w H^0_a(w)}{z-w},\\
      T(z)D(w) \sim \frac{D(w)}{(z-w)^2}+\frac{\partial_w D(w)}{z-w}.
\end{gathered}
\end{equation}
The mixed OPE between $H^{1}_{\pm \frac12}$ and $G^{-}_{\pm \frac12}$ takes the following general form, dictated by conformal invariance:
\begin{equation}
\begin{gathered}
   H^{1}_{\alpha}(z)G^{-}_{\beta}(w)=\epsilon_{\alpha\beta}\left(\frac{d_1}{(z-w)^3}+\frac{d_2~T(w)}{(z-w)}+\frac{d_3 \,\Xi(w)}{z-w}+\textcolor{black}{\frac{d_6\, \Lambda(w)}{z-w}+\frac{d_5\, \partial_w D(w)}{z-w}}\right)\\[6pt]
    +\left(\frac{2\,d_4\, (\lambda^a)_{\alpha\beta}\,H^{0}_a(w)}{(z-w)^2}+\frac{d_4\, (\lambda^a)_{\alpha\beta}\, \partial_w H^{0}_a(w)}{z-w}+\textcolor{black}{\frac{d_7~\Sigma(w)}{z-w}}\right)
    +\textcolor{black}{\frac{2\, d_5\, D(w)}{(z-w)^2}\epsilon_{\alpha\beta}},
\label{ope_two}
\end{gathered}
\end{equation}
where, \(\Xi(z)\), \(\Lambda(z)\), and \(\Sigma(z)\) are the quasi-primary operators defined as
\begin{equation}
\begin{gathered}
    \Xi(z):=\eta^{ab}(H^0_a\,H^0_b)(z),\quad \Lambda(z):=(D\,D)(z),\quad \Sigma(z):=(\lambda^a)_{\alpha\beta}(D\,H^0_a)(z).
\end{gathered}
\end{equation}
Here, \( (\lambda_{a})^{\alpha}_{~\beta}\)\,, \(\eta_{ab}\) and \(f^{c}_{~ab}\) are given in the main text \eqref{current_alg_coeff}. \(\epsilon_{\alpha\beta}\) is a antisymmetric tensor with \(\epsilon_{\frac12,-\frac12}=1\) and \((\lambda^a)_{\alpha\beta}=(\lambda_{b})^{\gamma}_{~\beta}~\eta^{ab}\,\epsilon_{\alpha\gamma}\).

To determine the coefficients \(q\)\,, Virasoro central charge \(c\) and \(d_1\) through \(d_7\)  in  \eqref{ope_two}, we impose the associativity condition on these OPEs. The resulting algebraic constraints can be efficiently solved using \texttt{Mathematica}~\cite{Thielemans:1991uw}, yielding
\begin{equation}
\label{coeff}
\begin{aligned}
    d_1 &= -\tfrac{1}{2}d_3(1+k), \quad
    d_2 =  \tfrac{1}{2}d_3(3+k), \quad
    d_4 =  d_3(1+k), \quad
    d_5 = -\tfrac{1}{2}d_3 k\sqrt{1+k}, \\[4pt]
    d_6 &= -\tfrac{3}{4}d_3(1+k), \quad
    d_7 =  2d_3\sqrt{1+k}, \quad
    q   =  \frac{1}{\sqrt{1+k}}, \quad
    c   =  \frac{3 + 3k - 6k^2}{3 + k},
\end{aligned}
\end{equation}

\subsubsection*{Free field representation}
\noindent For the free field representation of the above algebra, we will follow the construction of the papers by de Boer \emph{et al.} \cite{deBoer:1992sy,deBoer:1993iz} and also \cite{Bershadsky:1989mf}.
\begin{align}
\label{stensor}
    T(z)&=
-\frac{1}{12(4+\kappa)}\Bigg[
  6(4+\kappa)\,(\partial\phi_1\partial\phi_1)(z)
+ 6(4+\kappa)\,(\partial\phi_2\partial\phi_2)(z)
+ \big(24+6\kappa\big)\,(\partial\phi_3\partial\phi_3)(z)\nonumber\\
&\qquad
+ \big(48+12\kappa\big)\,\big(\beta_1\partial\gamma_1\big)(z)
+ \big(24+6\kappa\big)\,\big(\beta_2\partial\gamma_2\big)(z)
+ \big(24+6\kappa\big)\,\big(\beta_3\partial\gamma_3\big)(z)\nonumber\\
&\qquad- \big(24+6\kappa\big)\,\big(\partial\beta_2\,\gamma_2\big)(z)
- \big(24+6\kappa\big)\,\big(\partial\beta_3\,\gamma_3\big)(z)
- 3\,i\sqrt{2}\,\kappa\sqrt{4+\kappa}\;\partial^2\phi_1(z)\nonumber\\
&\qquad
+ i\sqrt{6}\,\sqrt{4+\kappa}\,(4+\kappa)\;\partial^2\phi_2(z)
+ 4\sqrt{3}\,\sqrt{-4-\kappa}\,(1+\kappa)\;\partial^2\phi_3(z)
\Bigg].
\end{align}
Note that the above \(T(z)\) has a central charge \(c=-\frac{3\kappa(3+2\kappa)}{4+\kappa}\). For the level matching, we have to replace \(k\) in \eqref{coeff} as \(k\rightarrow \kappa+1\).
\begin{align*}
D(z)=&-\frac{1}{\sqrt{\kappa+2}}\Bigg[ \frac{i\sqrt{4+\kappa}}{\sqrt{2}}\,\partial\phi_1(z)
- \frac{i\sqrt{4+\kappa}}{\sqrt{6}}\,\partial\phi_2(z)
+ \frac{i\sqrt{4+\kappa}}{\sqrt{3}}\,\partial\phi_3(z) \nonumber \\
& ~~~~~~~~~~~~~~~~~~~~~~~~~~~~~~~~~~~~~~~~~~~~~~~~~~~~~~~~~ + (\beta_2\gamma_2)(z)
+ (\beta_3\gamma_3)(z) \Bigg],\nonumber\\[6pt]
H^0_0(z)=&\frac{1}{2}\Big[
-4\,(\beta_1\gamma_1)(z)
+ 2\,(\beta_2\gamma_2)(z) 
-\, i\big(
   \sqrt{2}\sqrt{4+\kappa}\,\partial\phi_1(z) \nonumber\\
 & ~~~~~~~~~~~~~~~~~~~~~ + \sqrt{6}\sqrt{4+\kappa}\,\partial\phi_2(z)
 - 2i\,(\beta_3\gamma_3)(z)
\big)\Big],\nonumber\\[6pt]
H^0_{+1}(z)=&\beta_1(z),\nonumber\\[6pt]
H^0_{-1}(z)=&\frac{i\sqrt{4+\kappa}}{\sqrt{2}}\,
   (\partial\phi_1\,\gamma_1)(z)
\;+\;
i\sqrt{\frac{3}{2}}\sqrt{4+\kappa}\,
   (\partial\phi_2\,\gamma_1)(z)
\;+\;
\big(\beta_1(\gamma_1\gamma_1)\big)(z)\nonumber \\
&\quad
-\big(\beta_2(\gamma_1\gamma_2)\big)(z)
\;-\;(\beta_2\gamma_3)(z)
\;+\;
\big(\beta_3(\gamma_1\gamma_3)\big)(z)
\;+\;
\partial\gamma_1(z)
\;+\;
\kappa\,\partial\gamma_1(z),
\end{align*}
\begin{align}
G^{-}_{\frac{1}{2}}(z)=&-\frac{i\sqrt{4+\kappa}}{\sqrt{2}}\,
   (\partial\phi_1\,\beta_3)(z)
\;-\;
\frac{i\sqrt{4+\kappa}}{\sqrt{6}}\,
   (\partial\phi_2\,\beta_3)(z)
\;-\;
\frac{2i\sqrt{4+\kappa}}{\sqrt{3}}\,
   (\partial\phi_3\,\beta_3)(z)\nonumber\\
&\quad
-(\beta_1\beta_2)(z)
\;-\;
\big(\beta_3(\beta_3\gamma_3)\big)(z)
\;+\;
2\,\partial\beta_3(z)
\;+\;
\kappa\,\partial\beta_3(z)\,,\nonumber\\[8pt]
G^{-}_{-\frac{1}{2}}(z)=&\frac{i\sqrt{4+\kappa}}{\sqrt{2}}\big(\partial\phi_1(\beta_3\gamma_1)\big)(z)
+\frac{i\sqrt{4+\kappa}}{\sqrt{6}}\big(\partial\phi_2(\beta_3\gamma_1)\big)(z)
+i\sqrt{\tfrac{2}{3}}\sqrt{4+\kappa}\big(\partial\phi_2\beta_2\big)(z)\nonumber\\[-4pt]
&+\frac{2i\sqrt{4+\kappa}}{\sqrt{3}}\big(\partial\phi_3(\beta_3\gamma_1)\big)(z)
-\frac{2i\sqrt{4+\kappa}}{\sqrt{3}}\big(\partial\phi_3\beta_2\big)(z)
+\big(\beta_1(\beta_2\gamma_1)\big)(z)\nonumber\\[-4pt]
& -\big(\beta_2(\beta_2\gamma_2)\big)(z) -\big(\beta_2(\beta_3\gamma_3)\big)(z)
+\big(\beta_3(\beta_3(\gamma_1\gamma_3))\big)(z)
-2\big(\partial\beta_3\,\gamma_1\big)(z) \nonumber \\
& -\kappa\big(\partial\beta_3\,\gamma_1\big)(z)
+\partial\beta_2(z)
+\kappa\,\partial\beta_2(z)\,,
\end{align}
\begin{align}
H^{1}_{\frac{1}{2}}(z)=&-\frac{i\sqrt{4+\kappa}}{\sqrt{2}}\big(\partial\phi_1\,\gamma_2\big)(z)
+ i\sqrt{\tfrac{3}{2}}\sqrt{4+\kappa}\,\big(\partial\phi_2\,\gamma_2\big)(z)- (\beta_1\gamma_3)(z)\nonumber
\\[-4pt]
&- \big(\beta_2(\gamma_2\gamma_2)\big)(z)
- 2\,\partial\gamma_2(z)
- \kappa\,\partial\gamma_2(z)\,,\nonumber\\[6pt]
H^{1}_{-\frac{1}{2}}(z)=&\frac{i\sqrt{4+\kappa}}{\sqrt{2}}\big(\partial\phi_1(\gamma_1\gamma_2)\big)(z)
+ i\sqrt{2}\sqrt{4+\kappa}\,\big(\partial\phi_1\,\gamma_3\big)(z)
- i\sqrt{\tfrac{3}{2}}\sqrt{4+\kappa}\,\big(\partial\phi_2(\gamma_1\gamma_2)\big)(z)\nonumber
\\[-4pt]
&+\big(\beta_1(\gamma_1\gamma_3)\big)(z)
+\big(\beta_2(\gamma_1(\gamma_2\gamma_2))\big)(z)
+\big(\beta_2(\gamma_2\gamma_3)\big)(z)
+\big(\beta_3(\gamma_3\gamma_3)\big)(z)\nonumber
\\[4pt]
&+2\,\big(\gamma_1\,\partial\gamma_2\big)(z)
+\kappa\,\big(\gamma_1\,\partial\gamma_2\big)(z)
+\partial\gamma_3(z)
+\kappa\,\partial\gamma_3(z),
\end{align}
%

%
\section{Details of \texorpdfstring{$\mathcal{O}(\bar{w}^2)$}{O(z2)} 
terms in the graviton-graviton OPE}
\label{order_o2}
%
In this appendix, we collect the complete set of vertex operator OPEs needed to 
evaluate the $\mathcal{O}(\bar{w}^2)$ contribution to the graviton-graviton OPE. 
The $U$-$V$ OPEs, which contribute simple poles to the graviton primary exchange 
channel, are
\paragraph{The mixed $U$-$V$ OPEs.}
\begin{align*}
U_{h_1,\bar{h}_1+2}(z)\,V_{h_2,\bar{h}_2}(w)
&\sim \phantom{-}\frac{2\,\nu_2\,(\Delta_1-2)\,(\Delta_1-1)}
{\Delta_1+\Delta_2-1}
\frac{U_{h_1+h_2-1,\bar{h}_1+\bar{h}_2+2}(w)}{z-w}\,,
\\[4pt]
V_{h_1,\bar{h}_1}(z)\,U_{h_2,\bar{h}_2+2}(w)
&\sim -\frac{2\nu_2\,(\Delta_2-2)\,(\Delta_2-1)}
{\Delta_1+\Delta_2-1}
\frac{U_{h_1+h_2-1,\bar{h}_1+\bar{h}_2+2}(w)}{z-w}\,,
\\[4pt]
V_{h_1,\bar{h}_1+2}(z)\,U_{h_2,\bar{h}_2}(w)
&\sim \phantom{-}\frac{(\Delta_1-1)\,(\Delta_2-2)}
{\Delta_1+\Delta_2-2}
\frac{U_{h_1+h_2-1,\bar{h}_1+\bar{h}_2+2}(w)}{z-w}\,,
\\[4pt]
U_{h_1,\bar{h}_1}(z)\,V_{h_2,\bar{h}_2+2}(w)
&\sim -\frac{(\Delta_1-2)\,(\Delta_2-1)}
{\Delta_1+\Delta_2-2}
\frac{U_{h_1+h_2-1,\bar{h}_1+\bar{h}_2+2}(w)}{z-w}\,,
\\[4pt]
V_{h_1,\bar{h}_1+1}(z)\,U_{h_2,\bar{h}_2+1}(w)
&\sim \phantom{-}\frac{(\Delta_2-2)\,(\Delta_2-\Delta_1)}
{\Delta_1+\Delta_2-2}
\frac{U_{h_1+h_2-1,\bar{h}_1+\bar{h}_2+2}(w)}{z-w}\,,
\\[4pt]
U_{h_1,\bar{h}_1+1}(z)\,V_{h_2,\bar{h}_2+1}(w)
&\sim -\frac{(\Delta_1-2)\,(\Delta_1-\Delta_2)}
{\Delta_1+\Delta_2-2}
\frac{U_{h_1+h_2-1,\bar{h}_1+\bar{h}_2+2}(w)}{z-w}\,.
\end{align*}
\paragraph{The $V$-$V$ OPEs.} The $V$-$V$ OPEs generate both simple and double poles, with the double-pole 
coefficient being the scalar primary $\Phi_\Delta$:
\begin{equation*}
\begin{gathered}
V_{h_1,\bar{h}_1+1}(z)V_{h_2,\bar{h}_2+1}(w)
=-(\Delta_1-\Delta_2)\frac{V_{h_1+h_2-1,\bar{h}_1+\bar{h}_2+2}(w)}{(z-w)}
-2\,\alpha\,\frac{(\mathcal{V}_1\mathcal{V}_2)(w)}{(z-w)^2}
+\cdots\,,\\
V_{h_1,\bar{h}_1+2}(z)V_{h_2,\bar{h}_2}(w)
=(\Delta_1-1)\frac{V_{h_1+h_2-1,\bar{h}_1+\bar{h}_2+2}(w)}{(z-w)}
+\alpha\,\frac{(\mathcal{V}_1\mathcal{V}_2)(w)}{(z-w)^2}
+\cdots\,,\\
V_{h_1,\bar{h}_1}(z)V_{h_2,\bar{h}_2+2}(w)
=-(\Delta_2-1)\frac{V_{h_1+h_2-1,\bar{h}_1+\bar{h}_2+2}(w)}{(z-w)}
+\alpha\,\frac{(\mathcal{V}_1\mathcal{V}_2)(w)}{(z-w)^2}
+\cdots\,,
\end{gathered}
\end{equation*}
where, the ellipses denote descendant contributions,$\alpha = -1 + \Delta_1(\Delta_2-1) - \Delta_2 - \kappa$\\
\noindent and $(\mathcal{V}_1\mathcal{V}_2)(z)
= e^{i(r_2+s_2)(\sqrt{3}\,\phi_1+\phi_2-2i\,\phi_3)}(z)$.
%
%
\section{Details of \texorpdfstring{$\mathcal{O}(\bar{w}^3)$}{O(z3)} 
terms in the graviton-graviton OPE}
\label{order_o3}
%
\noindent In this appendix, we collect the complete set of vertex operator OPEs required to 
evaluate the $\mathcal{O}(\bar{w}^3)$ contribution to the graviton-graviton OPE.
\paragraph{The $V$-$V$ OPEs.}
\begin{align*}
V_{h_1,\bar{h}_1+2}(z)\,V_{h_2,\bar{h}_2+1}(w)
&= -\frac{(\Delta_1-1)(2+\Delta_1-2\Delta_2)}{-1+\Delta_1+\Delta_2}
\frac{V_{h_1+h_2-1,\bar{h}_1+\bar{h}_2+3}(w)}{z-w} \notag\\
&\quad -\,(2\Delta_1-\Delta_2)\bigl(-1+\Delta_1(\Delta_2-1)-\Delta_2-\kappa\bigr)
\frac{(\gamma_1\mathcal{V}_1\mathcal{V}_2)(w)}{(z-w)^2}
+\cdots\,,\\[8pt]
V_{h_1,\bar{h}_1+1}(z)\,V_{h_2,\bar{h}_2+2}(w)
&= \phantom{-}\frac{(\Delta_2-1)(2-2\Delta_1+\Delta_2)}{-1+\Delta_1+\Delta_2}
\frac{V_{h_1+h_2-1,\bar{h}_1+\bar{h}_2+3}(w)}{z-w} \notag\\
&\quad +\,(\Delta_1-2\Delta_2)\bigl(-1+\Delta_1(\Delta_2-1)-\Delta_2-\kappa\bigr)
\frac{(\gamma_1\mathcal{V}_1\mathcal{V}_2)(w)}{(z-w)^2}
+\cdots\,,\\[8pt]
V_{h_1,\bar{h}_1+3}(z)\,V_{h_2,\bar{h}_2}(w)
&= \phantom{-}\frac{3\,\mu_3\,(\Delta_1-1)\Delta_1}{1-\Delta_1-\Delta_2}
\frac{V_{h_1+h_2-1,\bar{h}_1+\bar{h}_2+3}(w)}{z-w} \notag\\
&\quad -\,3\,\mu_3\,\Delta_1\bigl(1+\Delta_1+\Delta_2-\Delta_1\Delta_2+\kappa\bigr)
\frac{(\gamma_1\mathcal{V}_1\mathcal{V}_2)(w)}{(z-w)^2}
+\cdots\,,\\[8pt]
V_{h_1,\bar{h}_1}(z)\,V_{h_2,\bar{h}_2+3}(w)
&= -\frac{3\,\mu_3\,(\Delta_2-1)\Delta_2}{1-\Delta_1-\Delta_2}
\frac{V_{h_1+h_2-1,\bar{h}_1+\bar{h}_2+3}(w)}{z-w} \notag\\
&\quad -\,3\,\mu_3\,\Delta_2\bigl(1+\Delta_1+\Delta_2-\Delta_1\Delta_2+\kappa\bigr)
\frac{(\gamma_1\mathcal{V}_1\mathcal{V}_2)(w)}{(z-w)^2}
+\cdots\,.
\end{align*}

\noindent
Again, the ellipses denote descendant contributions, and
\begin{equation*}
(\gamma_1\mathcal{V}_1\mathcal{V}_2)(z)
=
\left(\gamma_1\, e^{i(r_2+s_2)(\sqrt{3}\,\phi_1+\phi_2-2i\,\phi_3)}\right)(z)\,.
\end{equation*}
\paragraph{The $U$-$V$ OPEs.} The mixed sector at this order produces only simple 
poles, all proportional to $U_{h_1+h_2-1,\bar{h}_1+\bar{h}_2+3}$:
\begin{align*}
    U_{h_1,\bar{h}_1}(z)\,V_{h_2,\bar{h}_2+3}(w)
&\sim -\frac{3\mu_3\,(\Delta_1-2)\,(\Delta_2-1)\,\Delta_2}
{2\nu_2\,(\Delta_1+\Delta_2-2)\,(\Delta_1+\Delta_2-1)}
\frac{U_{h_1+h_2-1,\bar{h}_1+\bar{h}_2+3}(w)}{z-w}\,,
\\[4pt]
V_{h_1,\bar{h}_1+3}(z)\,U_{h_2,\bar{h}_2}(w)
&\sim \phantom{-}\frac{3\mu_3\,(\Delta_1-1)\,\Delta_1\,(\Delta_2-2)}
{2\nu_2\,(\Delta_1+\Delta_2-2)\,(\Delta_1+\Delta_2-1)}
\frac{U_{h_1+h_2-1,\bar{h}_1+\bar{h}_2+3}(w)}{z-w}\,,
\\[4pt]
U_{h_1,\bar{h}_1+1}(z)\,V_{h_2,\bar{h}_2+2}(w)
&\sim -\frac{(\Delta_1-2)\,(2\Delta_1-\Delta_2-2)\,(\Delta_2-1)}
{2\nu_2\,(\Delta_1+\Delta_2-2)\,(\Delta_1+\Delta_2-1)}
\frac{U_{h_1+h_2-1,\bar{h}_1+\bar{h}_2+3}(w)}{z-w}\,,
\\[4pt]
V_{h_1,\bar{h}_1+2}(z)\,U_{h_2,\bar{h}_2+1}(w)
&\sim -\frac{(\Delta_1-1)\,(\Delta_1-2\Delta_2+2)\,(\Delta_2-2)}
{2\nu_2\,(\Delta_1+\Delta_2-2)\,(\Delta_1+\Delta_2-1)}
\frac{U_{h_1+h_2-1,\bar{h}_1+\bar{h}_2+3}(w)}{z-w}\,,
\end{align*}
\begin{align*}
U_{h_1,\bar{h}_1+2}(z)\,V_{h_2,\bar{h}_2+1}(w)
&\sim -\frac{(\Delta_1-2)\,(\Delta_1-1)\,(\Delta_1-2\Delta_2+2)}
{(\Delta_1+\Delta_2-2)\,(\Delta_1+\Delta_2-1)}
\frac{U_{h_1+h_2-1,\bar{h}_1+\bar{h}_2+3}(w)}{z-w}\,,
\\[4pt]
V_{h_1,\bar{h}_1+1}(z)\,U_{h_2,\bar{h}_2+2}(w)
&\sim \phantom{-}\frac{(\Delta_2-2)\,(\Delta_2-1)\,(\Delta_2-2\Delta_1+2)}
{(\Delta_1+\Delta_2-2)\,(\Delta_1+\Delta_2-1)}
\frac{U_{h_1+h_2-1,\bar{h}_1+\bar{h}_2+3}(w)}{z-w}\,,
\\[4pt]
U_{h_1,\bar{h}_1+3}(z)\,V_{h_2,\bar{h}_2}(w)
&\sim \phantom{-}\frac{3\nu_3\,(\Delta_1-2)\,(\Delta_1-1)\,\Delta_1}
{(\Delta_1+\Delta_2-2)\,(\Delta_1+\Delta_2-1)}
\frac{U_{h_1+h_2-1,\bar{h}_1+\bar{h}_2+3}(w)}{z-w}\,,
\\[4pt]
V_{h_1,\bar{h}_1}(z)\,U_{h_2,\bar{h}_2+3}(w)
&\sim -\frac{3\nu_3\,(\Delta_2-2)\,(\Delta_2-1)\,\Delta_2}
{(\Delta_1+\Delta_2-2)\,(\Delta_1+\Delta_2-1)}
\frac{U_{h_1+h_2-1,\bar{h}_1+\bar{h}_2+3}(w)}{z-w}\,.
\end{align*}
%
%
\section{Symmetries in conformal gravity}\label{symmetry_algebra}
%

In \cite{Ghorai:2025ebc}, we computed the graviton-graviton OPE on the celestial sphere from the bulk MHV amplitudes of the BW theory. It is given by,
\begin{equation}\label{gg_OPE_summarys}
\begin{gathered}
G^{++}_{\Delta_1}(z,\bar z)\, G^{++}_{\Delta_2}(w,\bar w) = - \frac{(\bar z - \bar w)}{(z-w)}\,B(\Delta_1-1,\Delta_2-1)\,G^{++}_{\Delta_1+\Delta_2}(w,\bar w) \\[4pt]
 - \frac{(\bar z - \bar w)^2}{(z-w)}\,B(\Delta_1,\Delta_2-1)\, \partial_{\bar w} G^{++}_{\Delta_1+\Delta_2}(w,\bar w) 
 - \frac{(\bar z- \bar w)^2}{(z-w)^2}\, B(\Delta_1,\Delta_2)\, \Phi_{\Delta_1+\Delta_2}(w,\bar w) \\[4pt]
 - \frac{(\bar z - \bar w)^2}{(z-w)}\, B(\Delta_1+1,\Delta_2)\, \partial_{w}\Phi_{\Delta_1+\Delta_2}(w,\bar w) + \cdots
 \end{gathered}
\end{equation}

We now show that the $\mathfrak{sl}(2,\mathbb{R})$ current algebra consistent with the above OPE has a central extension. 
Invoking the soft-limit relation in ~\eqref{soft-limits}, we obtain the three $\mathfrak{sl}(2,\mathbb{R})$ current--graviton OPEs:
\begin{equation}
\label{sl_currents_OPEs}
\begin{gathered}
H^0_1(z)\, G^{++}_{\Delta}(w,\bar w) \sim  
- \frac{(\Delta-2)\,\bar w}{(z-w)}\,G^{++}_\Delta(w, \bar w)
- \frac{\bar w^2}{(z-w)}\, \partial_{\bar w}G^{++}_\Delta(w, \bar w)
- \frac{\bar w^2}{(z-w)^2}\, \Phi_\Delta(w, \bar w)\,, \\[4pt]
H^0_0(z)\, G^{++}_{\Delta}(w,\bar w) \sim 
\frac{(\Delta-2)}{(z-w)}\,G^{++}_\Delta(w, \bar w)
+ \frac{2\bar w}{(z-w)}\,\partial_{\bar w} G^{++}_\Delta(w, \bar w)
+ \frac{2\bar w}{(z-w)^2}\, \Phi_\Delta(w,\bar w)\,, \\[4pt]
H^0_{-1}(z)\, G^{++}_{\Delta}(w,\bar w) \sim 
-\frac{1}{(z-w)}\,\partial_{\bar w} G^{++}_\Delta(w, \bar w)
- \frac{1}{(z-w)^2}\, \Phi_\Delta(w,\bar w)\,.
\end{gathered}
\end{equation}
A further soft limit now reveals the current-current OPE. Taking the subleading soft limit 
of the second relation in \eqref{sl_currents_OPEs}, expanding both $G^{++}_{\Delta}(w,\bar{w})$ 
and $\Phi_{\Delta}(w,\bar{w})$ in terms of soft currents, and isolating the term linear in  $\bar{w}$, we find
\begin{equation}
H^{0}_{0}(z)\,H^{0}_{0}(w) \sim -\frac{2}{(z-w)^2}\,.
\end{equation}
This demonstrates that the $\mathfrak{sl}(2,\mathbb{R})$ algebra acquires a non-vanishing 
central extension in the BW theory. Strikingly, this result is in precise agreement with 
the  free-field realisation of the $\mathfrak{sl}(2,\mathbb{R})_{\kappa}$ boundary symmetry algebra ~\eqref{chir_curr} 
evaluated at level $\kappa = -2$,
\begin{equation}\label{central_term_current_algebra}
H^{0}_{0}(z)\,H^{0}_{0}(w) \sim \frac{2(\kappa+1)}{(z-w)^2}\bigg|_{\kappa=-2} = -\frac{2}{(z-w)^2}\,.
\end{equation}
The remaining $\mathfrak{sl}(2,\mathbb{R})_\kappa$ current-current OPEs can be worked out 
order by order in $\bar{w}$, and in each case one finds exact agreement with 
~\eqref{ope_one} at level $\kappa = -2$.

%
%
\bibliographystyle{}
\providecommand{\href}[2]{#2}\begingroup\raggedright

\end{document}